\documentclass[traditabstract]{aa} 
\usepackage{graphicx}
\usepackage{equationarray}
\usepackage{xcolor,soul}
\usepackage{multirow}
\usepackage{booktabs}
\usepackage{nicefrac}
\usepackage{amsmath}

\usepackage{txfonts}
\usepackage{natbib}
\newcommand{\nc}{\newcommand}
\nc{\lsun}{\ensuremath{\mathrm{L}_\odot}}
\nc{\msun}{\ensuremath{\mathrm{M}_\odot}}
\nc{\tex}{\ensuremath{\mathrm{T}_{\rm ex}}}
\nc{\cthree}{C$_3$}
\nc{\ctwo}{C$_2$}
\nc{\thCCC}{$^{13}$CCC}
\nc{\CthCC}{C$^{13}$CC}
\nc{\cplus}{C$^+$}
\nc{\kms}{\mbox{km\,s$^{-1}$}}
\nc{\Kkms}{\mbox{K\,km\,s$^{-1}$}}
\nc\micron{\mbox{$\mu$m}}
\nc{\Trot}{T$_{\rm rot}$}%
\nc{\Ntot}{N$_{\rm tot}$}%
\nc{\Tbg}{T$_{\rm bg}$}%
\nc{\cmcub}{\mbox{cm$^{-3}$}}
\nc{\cmsq}{\mbox{cm$^{-2}$}}
\newcommand\arcdeg{\mbox{$^\circ$}}%

\begin{document}
\title{First detection of  the carbon chain molecules \thCCC\  and
\CthCC\  towards SgrB2(M)}
   \author{
T. F. Giesen\inst{\ref{kassel}}, 
B. Mookerjea\inst{\ref{tifr}}, 
G. W. Fuchs\inst{\ref{kassel}}, 
A.  A. Breier\inst{\ref{kassel}}, 
D. Witsch\inst{\ref{kassel}}, 
R. Simon\inst{\ref{kosma}}, 
J. Stutzki\inst{\ref{kosma}}, 
}

\institute{Laboratory for Astrophysics, Institute of Physics, University
of Kassel, Kassel D-34132, Germany
\email{t.giesen@uni-kassel.de}\label{kassel}
\and
Tata Institute of Fundamental Research, Homi Bhabha Road,
Mumbai 400005, India \label{tifr} 
\and 
I. Physikalisches Institut, University of Cologne, Germany\label{kosma}
}

 
\date{Received \ldots accepted \ldots}

 \abstract
{
}

\keywords{ISM:~molecules -- Submillimeter:~ISM -- Molecular processes}
 
\titlerunning{$^{13}$CCC \& C$^{13}$CC isotopologues towards SgrB2(M)}
        \authorrunning{Giesen et al.}

\abstract 
{
{\it Context.}  Carbon molecules and their $^{13}$C-isotopologues can be used to determine the $^{12}$C/$^{13}$C abundance ratios in stellar and interstellar objects. C$_3$ is a pure carbon chain molecule found in star forming regions and in stellar shells of carbon-rich late-type stars. Latest laboratory data of $^{13}$C-isotopologues of C$_3$ allow a selective search for the mono-substituted species $^{13}$CCC and C$^{13}$CC based on accurate ro-vibrational frequencies.\\ 
{\it Aims.} Our aim was to provide the first detection of the  $^{13}$C-isotopologues \thCCC\, and \CthCC\, in space and to derive the $^{12}$C/$^{13}$C ratio of interstellar gas in the massive star-forming region SgrB2(M) near the Galactic Center. \\ 
{\it Methods.} We used the heterodyne receivers GREAT and upGREAT on board SOFIA to search for the ro-vibrational transitions $Q$(2) and $Q$(4) of  \thCCC\, and \CthCC\, at 1.9 THz along the line of sight towards
SgrB2(M). In addition, to determine the local excitation temperature we analyzed data from nine ro-vibrational transitions of the main isotopologue CCC in the frequency range between 1.6 - 1.9 THz which were taken from the Herschel Science Data Archive.\\  
{\it Results. } We report the first detection of the isotopologues \thCCC\, and
\CthCC . For both species the ro-vibrational absorption lines $Q$(2) and $Q$(4)
have been identified, primarily arising from the warm gas physically associated
with the strong continuum source SgrB2(M). From the available CCC
ro-vibrational transitions we derived a gas excitation temperature of $T_{ex}$
= 44.4$^{+4.7}_{-3.9}$~K and a total column density of
$N$(CCC)=3.88$^{+0.39}_{-0.35}\times 10^{15}$\,\cmsq .  Assuming the excitation
temperatures of \CthCC\ and \thCCC\  to be the same as for CCC, we obtained
column densities of the $^{13}$C-isotopologues of $N$(\CthCC) =
2.1$^{+0.9}_{-0.6}\times 10^{14}$\,\cmsq\  and
$N$(\thCCC)=2.4$^{+1.2}_{-0.8}\times 10^{14}$\,\cmsq.  The derived
$^{12}$C/$^{13}$C abundance ratio in the C$_3$ molecules is  20.5$\pm
4.2$,  which is in agreement with the elemental ratio of 20, typically observed
in SgrB2(M).  However, we find the $N$(\thCCC) / $N$(\CthCC)  ratio to be
1.2$\pm$0.1, which is shifted from the statistically expected value of 2. We
propose that the discrepant abundance ratio arises due to the lower
zero-point energy of \CthCC\ which makes position-exchange reaction
converting \thCCC\  to \CthCC\ energetically favorable.}

\maketitle

\section{Introduction}

Small carbon chain molecules play an important role in the
chemistry of stellar and interstellar environments since they are
ubiquitous throughout the interstellar medium \citep{adamkovics2003} and
most likely participate in the formation of long carbon chain
molecules.  Furthermore, they are products of photo-fragmentation
cascades of polycyclic aromatic hydrocarbons (PAHs)
\citep{radi1988,pety2005}.  \citet{oka2003} observed the \cthree\ column
densities to be well correlated with the corresponding C$_2$ column
densities along several translucent sight lines and concluded that
\cthree\ and \ctwo\ are formed in the same chain of chemical reactions.
However, this chemical pathway can not be used to explain the abundance
of \cthree\ in the warm envelopes of hot cores. Post warm-up gas phase
chemistry of CH$_4$ applicable at moderate temperatures of 30\,K  is
required to explain the abundance of \cthree\ and associated hydrocarbons
satisfactorily for a few sources
\citep{mookerjea2010,mookerjea2012,mookerjea2014}.  Further observations
of high density warm regions are needed to clearly establish this
hypothesis of carbon chain formation in these regions.

Studies of isotope fractionation  have been proven to be a useful tool
of tracing chemical reaction pathways and to elucidate formation and
destruction processes of interstellar molecules. At low temperatures the
isotopic ratio of molecular carbon can be significantly shifted due to
small zero-point energy differences between reactants and products
\citep[e.g.][]{langer1984}.  \citet{takano1998} were among the first to
observe the three singly substituted  $^{13}$C-species of HC$_3$N and
found that the abundance of HCC$^{13}$CN is significantly higher than
those for H$^{13}$CCCN and HC$^{13}$CCN, thus indicating that the three
carbon atoms are not equivalent in the formation pathways of HC$_3$N.
Similar results have also been observed for CCS, CCH, C$_3$S, HC$_3$N,
 $c$-C$_3$H$_2$
\citep{sakai2007,sakai2010,sakai2013,taniguchi2016a,yoshida2015}, whereas for
HC$_5$N no significant difference in the abundance of the five singly
substituted isotopologues was observed \citep{taniguchi2016b}.  The
observed anomalies in isotopic ratios have been explained by single
isotope-specific reactions in the formation pathways. Based on a
gas-grain chemical network, \citet{furuya2011} confirmed that the
isotope ratios of molecules, both in the gas-phase and on grain
surfaces, mostly depend on whether these species are formed from the
carbon atom (ion) or the CO molecule.  These authors concluded that the
$^{12}$C/$^{13}$C isotope ratio is large if the species is formed from a
carbon atom, while the ratio is small if the species is formed from a CO
molecule. While the evidence for isotope-specific fractionation of
carbon is strongest in low-mass star forming regions, a few instances
are also seen in high mass cores such as G28.28-0.36
\citep{taniguchi2016a}. 

The Sagittarius B2 molecular cloud (Sgr\,B2) is a very massive (a
few 10$^6$~\msun) and extremely active region of high-mass star
formation with an extraordinarily rich chemistry. It is located at a
projected distance of 107 pc from Sgr\,A$^\ast$, the compact radio
source at the Galactic center and  (8.34$\pm$0.16) kpc away from the Sun
\citep{Reid2009}. Sgr\,B2 contains at least two main sites of star
formation, Sgr\,B2(N) ($\upsilon$ = 64\,\kms) and Sgr\,B2(M) ($\upsilon$
= 62\,\kms), which are separated by 48\arcsec. Both these sources have
proven to be extremely fertile hunting grounds for complex organic
molecules \citep{belloche2013}. Owing to its strategic location and
strong sub-millimeter continuum flux Sgr\,B2 is one of the best
suited background sources towards which absorption studies can be carried out.
Because of its higher continuum temperature \citep{Qin2008}, we selected Sgr\,B2(M) as the background source for the detection of
absorption lines of \cthree\ and its singly substituted $^{13}$C-isotopologues.

\section{Spectroscopic Data}

In the remainder of the paper we use the term C$_3$ generally when the
triatomic carbon molecule (Tricarbon) is meant without considering the
specific isotopic composition, and we use the terms CCC, \CthCC\, and
\thCCC\, for the  isotopologues $^{12}$C$^{12}$C$^{12}$C,
$^{12}$C$^{13}$C$^{12}$C, and $^{13}$C$^{12}$C$^{12}$C, respectively.

C$_3$ is a floppy linear molecule of electronic ground state
$^1\Sigma_g^+$. As a centro-symmetric molecule, C$_3$ has no permanent
electric dipole moment and thus, it lacks a rotational spectrum, but a
vibrationally induced electric dipole moment allows for ro-vibrational
transitions of the asymmetric stretching $\nu_3(\sigma_u)$  and the
energetically low-lying bending mode $\nu_2(\pi_u)$.
\citet{matsumura1988} assigned rotationally resolved infrared spectra to
the $\nu_3(\sigma_u)$ fundamental band of the CCC main isotopologue at
2040 cm$^{-1}$ (61 THz).  Later, \citet{moazzen-ahmadi1993} reported
$\nu_3(\sigma_u)$ spectra of the \thCCC\ isotopologue. CCC spectra of
the energetically low-lying $\nu_2(\pi_u)$ mode at 63 cm$^{-1}$ (2~THz)
were published by \citet{schmuettenmaer1990} and \cite{gendriesch2003}.

Recently, \citet{Breier2016} measured the lowest bending mode, $\nu_2(\pi_u)$, of the five singly and multiply substituted $^{13}$C-isotoplologues of C$_3$.  Using these accurate frequencies we have undertaken
an observational study of CCC, \CthCC , and \thCCC\, to determine the
$^{12}$C/$^{13}$C isotopic ratio in the dense  warm molecular gas
along the line of sight to Sgr\,B2(M).

The $\nu_2(\pi_u)$ bending mode spectrum consists of $P$-, $R$- and strong
$Q$-branch transitions which are rotationally resolved.  In the present
analysis to derive C$_3$ column densities we used a vibrational dipole
moment ($\mu_v$) of 0.432\,D, which is based on the best currently
available dipole moment surface \citep{Schroeder2016}.  The new value of
$\mu_v$ is slightly smaller than the commonly used value of 0.437\,D
published by \citet{Jensen1992}. For the analysis of the observed CCC,
\CthCC\ and \thCCC\ lines we used the molecular parameters obtained from
our recent laboratory study \citep{Breier2016}, which also includes
relevant \cthree\,data from the literature. Relative to the main
isotopologue, CCC, the band centers of $^{13}$C-substituted species are
shifted toward lower frequencies and their rotational constants  are
slightly smaller than that of the main isotopologue.  We assume the
$\nu_2$-vibrational dipole moment of all C$_3$-isotopologues to be the
same, which is a reasonable approximation.

A noticeable difference between the spectra of \cthree-isotopologues is
caused by the nuclear spin ($I_{\rm C}$=0) of identical $^{12}$C atoms
located  at both ends of the carbon chain.   Due to spin statistical
weights only even-numbered $J$ rotational levels exist for the
centro-symmetric molecules CCC and \CthCC, whereas all $J$ rotational
states are present in the asymmetrically substituted \thCCC . Due to the
absence of every second rotational level, the values of the
ro-vibrational partition functions, $Q_{rv}$, of the centro-symmetric
species are half as large as that of the asymmetrically substituted
\thCCC . 

\begin{center} 
 $Q_{rv}^{\rm CCC} =  Q_{rv}^{\rm C^{13}CC} = 0.5 \cdot Q_{rv}^{^{13}\rm CCC} $
\end{center}
 
On the other hand, if we assume that  $^{13}$C is statistically distributed
in \cthree, we find two options to place $^{13}$C asymmetrically at the ends of the carbon chain, but only one option to substitute the central carbon atom. Thus, there are twice as many asymmetrically substituted \thCCC\, species   as symmetrically substituted \CthCC\, species, and their relative abundance is:
\begin{center}
[\CthCC]/[\thCCC] = 0.5.
\end{center}  
With regard to the ro-vibrational line intensities, the double excess of \thCCC\  is compensated by the twice as large partition function $Q_{rv}^{^{13}\rm CCC} $, and the line intensities of, e.g., $Q$(2) and $Q$(4) of  \thCCC\  and \CthCC\ are expected to be the same, provided that $^{13}$C is purely statistically distributed and the isotopologue abundances are not affected by any isotope specific mechanism. \\

We calculated the Einstein $A_{ul}$ coefficients in  {\it SI}-units  for
the observed transitions using \citep{Bernath1995}: 

\begin{equation} 
\, \, \, A_{ul} = \frac{16 \pi^3}{3 \epsilon_0 h}\frac{\nu^3}{c^3} \cdot
|\mu_{lu}|^2  {\rm \, \, \, \, \, \, and \, \, \, \, \, \,  }
|\mu_{lu}|^2=\frac{|\mu_v|^2L_{P/Q/R}(J)}{g_u}, 
\end{equation} 

where, $|\mu_{lu}|^2$ is the square of the transition moment matrix
element and $\mu_{v} $ the vibrational dipole moment.  $L_{P/Q/R}(J)$
are the H{\"o}nl-London factors for  $P$-, $Q$-, or $R$-type transitions for
absorption from a lower $J_l$  state, and  $g_u=2J_u+1$ is the
degeneracy of the upper state,  $J_u$.  For ro-vibrational transitions
of a linear molecule from ground state to an degenerated  first excited
bending state the H{\"o}nl-London factors are $L_P(J) = J-1, L_Q(J) =
2J+ 1$ and $L_R(J) = J +2$, as given by \citet{ Hansson2005}.  The
Einstein  $A_{ul}$ coefficients derived in this way are  consistent
with the results of the PGopher program \citep{Pgopher} for calculating
line spectra and with the formalism for line intensities as described by
\citet{Bunker2005}. Table\,\ref{tab_labdata} summarizes the
spectroscopic parameters for all observed transitions.

\begin{table}
\caption{Spectroscopic parameters for the observed CCC,  C$^{13}$CC and $^{13}$CCC  transitions.
\label{tab_labdata}}
\begin{tabular}{lccrrr}
\hline
\hline
Transition & Frequency & Einstein $A_{ul}$ & E$_{l}$ & $g_l$ & $g_u$ \\
           & MHz & s$^{-1}$ & K & &\\
\hline\\
CCC &&&&&\\ 
$Q$(2)    & 1890558.188 & 0.01468 &  3.7 &  5 &  5 \\
$Q$(4)    & 1896706.838 & 0.01482 & 12.4 &  9 &  9 \\
$Q$(6)    & 1906337.907 & 0.01505 & 26.0 & 13 & 13 \\
$P$(2)    & 1836823.502 & 0.00449 &  3.7 &  5 &  3 \\
$P$(4)    & 1787890.534 & 0.00532 & 12.4 &  9 &  7 \\
$P$(6)    & 1741122.646 & 0.00521 & 26.0 & 13 & 11 \\
$P$(8)    & 1696525.363 & 0.00495 & 44.6 & 17 & 15 \\
$P$(10)   & 1654087.900 & 0.00466 & 68.1 & 21 & 19 \\
$P$(12)   & 1613805.250 & 0.00437 & 96.6 & 25 & 23 \\
\hline
 &&&&&\\
C$^{13}$CC &&&&&\\
$Q$(2) & 1819596.013 & 0.01309 &  3.7 & 5 & 5 \\
$Q$(4) & 1825647.312 & 0.01322 & 12.4 & 9 & 9 \\
\hline
 &&&&&\\
$^{13}$CCC &&&&&\\
$Q$(2)& 1882638.269 & 0.01450 &  3.6 & 5 & 5 \\
$Q$(4)& 1888501.880 & 0.01463 & 11.9 & 9 & 9 \\
\hline
\hline
\end{tabular}
\end{table}

\section{Observations}

\subsection{SOFIA}

Observations of the $Q$(2) and $Q$(4) transitions of \thCCC\  and \CthCC\ were
performed with the GREAT instrument, in 2015 in the single-pixel
configuration \citep{heyminck2012} and in  2016 - 2017 using the upGREAT
array configuration \citep{Risacher16} onboard the Stratospheric
Observatory for Infrared Astronomy \citep[SOFIA;][]{young2012}.  The
observations of \thCCC\, were obtained during two flights of the SOFIA/GREAT flight campaigns in July 2015 and in June 2016 from New Zealand. Observations
of the$ Q$(2) and $Q$(4) transitions of \CthCC\ were performed with
upGREAT/SOFIA during the New Zealand campaign in June 2017.

All observations were made in double beam-switch mode, in which the
source emission is alternately placed in one of the two chopper beams,
while the other beam  points at an off-source positions on either side of the
source.  This observing mode cancels the systematics related to the
optical path, such as the differences in standing wave between the two
chopped beams. This compensation is particularly important given the
strong continuum offset of the  SgrB2(M) source. The source position
used for the main continuum peak on SgrB2, i.e. SgrB2(M), was
($\alpha_{2000}$,$\delta_{2000}$)=(17$^h$47$^m$20\fs16, -28\arcdeg
23\arcmin 04\farcs5). The chopper amplitude was set to 80\arcsec ,
resulting in a chopper-throw of 160\arcsec , which was aligned at an
angle rotated counter-clockwise by 30\arcdeg\ relative to the
R.\,A.\,-axis. The telescope pointing rms deviations, judged by the
optical images on SOFIA's instrument focal plane pointing camera, as
well as from the SOFIA housekeeping data, was well below 1\arcsec.

For the 2016 and 2017 observations we used the central pixel (PX\_00) of the upGREAT
low-frequency-arrays in both horizontal (H) and vertical (V)  polarization. The stability of the
V-polarization Local Oscillator (LO) in the June 2017 observations was marginal, resulting in an
inconsistent calibration of the spectra obtained in this polarization. We therefore discarded these
observations from further analysis. 
 
\begin{table}
\caption{\CthCC\, and \thCCC\ observational parameters for various spectral line tunings of the SOFIA/GREAT (G) and upGREAT (upG) receivers.
\label{tab_obs}}      
\begin{tabular}{rc r r r r r }      
\hline\hline 
Transition & Rec. & Obs. Date & $\upsilon^\ast_{\rm off}$ & $T_{\rm
int}$ & $T_{\rm rms}$\\
 & &  & km/s & min & mK\\
\hline
\thCCC\ - $Q$(2) & G & 19 July 2015& +10.0  &  6.8 & 99\\
\thCCC\ - $Q$(4) & G & 19 July 2015  & +0.0   & 27.9 & 62\\
           &    upG & 9 June 2016 & -10.0  & 12.5 & 96 \\ \ \\
\CthCC\ - $Q$(2) & upG & 28 June 2017 &   0.0  &  5.7 & 112\\
        &    upG  & 28 June 2017 & +10.0  &  5.1 & 150\\
\CthCC\ -  $Q$(4)&  upG  & 28 June 2017 &   0.0  &  8.5 & 56\\
             &   upG     & 28 June 2017       & +10.0  &  2.8 & 150\\
\hline   
\end{tabular}
{\small
$^\ast$ Velocity offset  of LO setting relative to the $\upsilon_{\rm LSR}$
}
\end{table}

In order to identify line contamination from the image side-band, the
observations were performed with different LO tuning offsets from the
nominal setting. Table\,\ref{tab_obs} lists the details of the
observations, including the total integration times, $T_{\rm int}$, and the rms
achieved.  In the June 2016 observations the  \thCCC\ $Q$(4) line in
particular was severely affected by line contamination from the image
side-band. The velocity profile of the  unknown and unassigned  
weak line from the
image side-band was re-constructed by shifting the observed spectra from
the three tunings of July 2015 observations to the correct
image-side-band velocity scale.  The  line profile of the line
contaminating the June 2016 \thCCC\ $Q$(4) observations was  thus
determined and subtracted from the latter. The corrected June 2016
spectrum was in good agreement with the $Q$(4) spectrum measured in the
three tunings during the July 2015 campaign, so that both spectra could be
averaged together, resulting in the final spectrum shown in
Figure\,\ref{fig_13c3specs}.

\subsection{SOFIA archives \label{sec_archives}}

We obtained observations of the CCC $Q$(6) transition from the SOFIA data archive. The original observations were part of the proposal 01\_151 (PI.: P. Schilke, D. Neufeld). The observations were performed with GREAT on July 17, 2013 using two LO settings corresponding to offsets of -15\,kms\ and -25\,\kms\  with a total integration time of 15\,minutes. The spectrum was smoothed to  a resolution of 0.2\,\kms\ and had an rms of 10~mK.

 \subsection{Herschel HEXOS data}

The guaranteed time key project HEXOS \citep[Herschel Observations
of Extra-Ordinary Sources]{Bergin2010} included a full spectral survey
of SgrB2(M). Here we have used the User Provided Data Product (UPDP)
for the HIFI bands 7a and 7b from the Herschel Science Archive in
order to study all transitions, except $Q$(6) of the CCC main
isotopologue, which was taken from the SOFIA archive (see
Sec.\,\ref{sec_archives}).

\section{Results \& Analysis}

\begin{figure}[h]
\begin{center}
\includegraphics[width=0.47\textwidth]{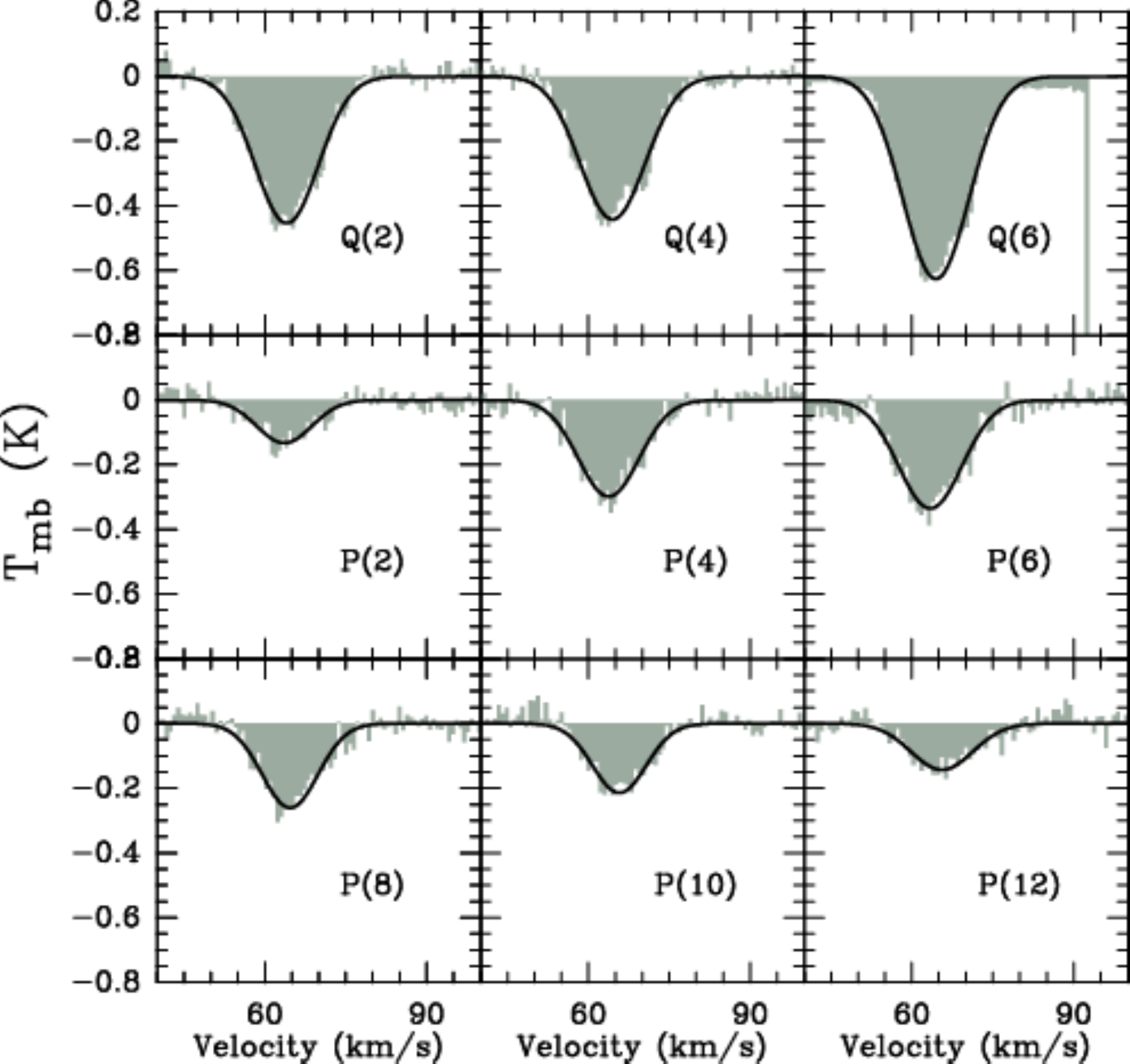}
\caption{Ro-vibrational spectra of CCC observed towards SgrB2(M) using
HIFI/Herschel along with fitted Gaussian profiles (smooth curve).
\label{fig_c3specs}}
\end{center}
\end{figure}

\begin{figure}
\begin{center}
\includegraphics[width=0.47\textwidth]{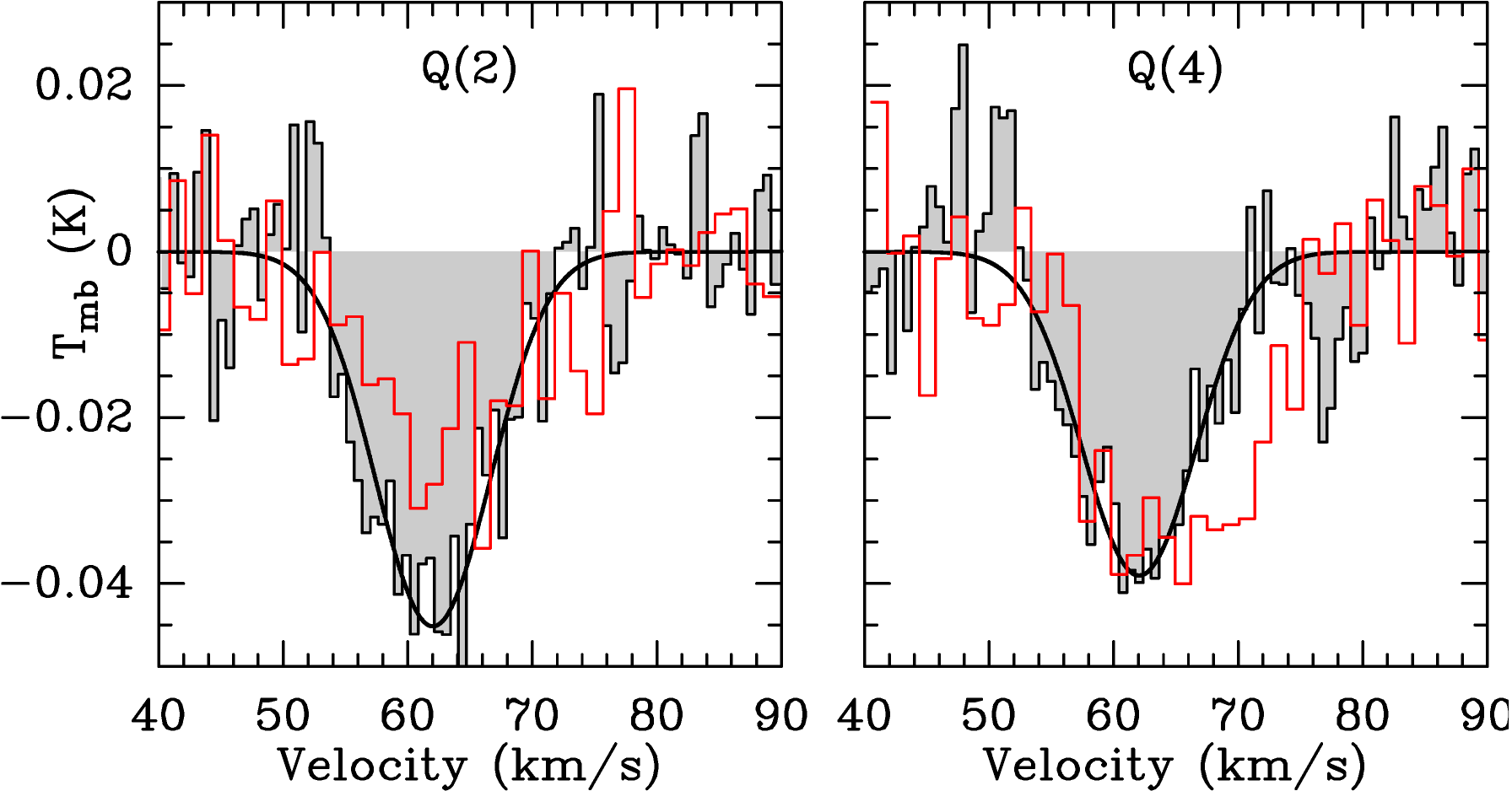}
\caption{Ro-vibrational spectra of C$^{13}$CC\ observed towards SgrB2(M)
using upGREAT/SOFIA along with fitted Gaussian profiles (smooth
curve). The H-polarization is shown as filled spectrum and the V-polarization as red spectrum.}
\label{fig_c13ccspecs}
\end{center}
\end{figure}

Figures\,\ref{fig_c3specs}\,--\,\ref{fig_c13ccspecs} show
the observed spectra of all the transitions of CCC, \thCCC\ and
\CthCC. Each spectrum was fitted with a single Gaussian component
   to obtain the velocities $\upsilon_{\rm LSR}$,  line widths $\Delta\upsilon$, and the integrated optical
depths (Table\,\ref{tab_obsdata}).

\begin{figure}[h]
\begin{center}
\includegraphics[width=0.44\textwidth]{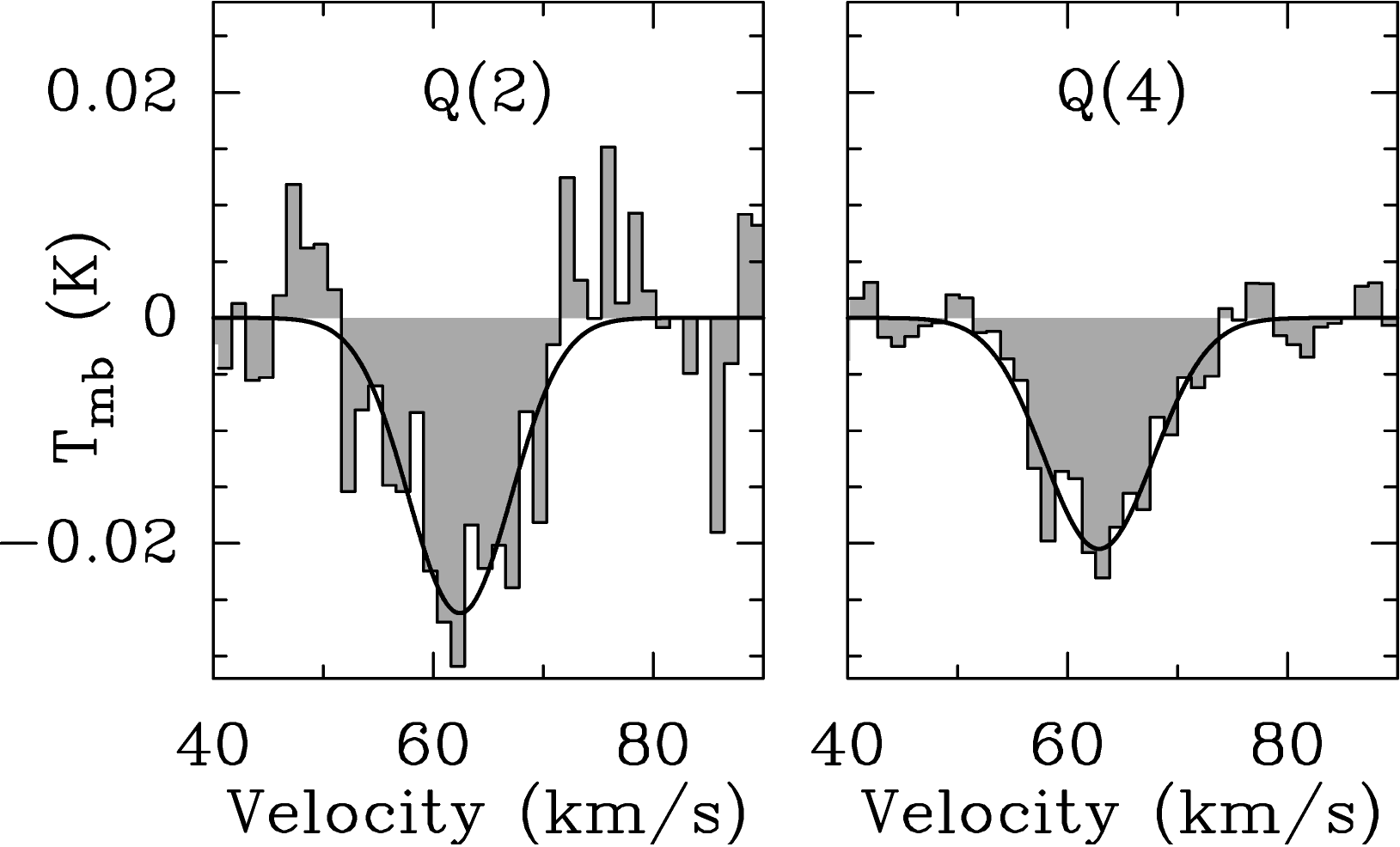}
\caption{Ro-vibrational spectra of $^{13}$CCC observed towards SgrB2(M)
using GREAT/SOFIA along with fitted Gaussian profiles (smooth curve).
\label{fig_13c3specs}}
\end{center}
\end{figure}

\begin{table}
\caption{Summary of observed and derived quantities for the CCC,   C$^{13}$CC, and
$^{13}$CCC  transitions.}
\label{tab_obsdata}
\begin{tabular}{llllrl}
\hline
\hline \vspace{-0.2cm}\\ 
 & $\tau_{\rm peak}$ & $\upsilon_{\rm LSR}$ & $\Delta \upsilon$ & $\int{\tau {\rm d} \upsilon}$\,\, \,& $N_{J_l}\times 10^{13}$\\
           &                   & \kms& \kms &\kms & \cmsq\\
\hline \vspace{-0.2cm}\\ 
\multicolumn{2}{l}{CCC}  &&&&\\
\vspace{-0.3cm}\\ 
$Q$(2)  & 0.53 & 63.9(1) & 11.6(2) &  6.6(1) & 32.6(5)\\
$ Q$(4)  & 0.56 & 64.4(1) & 13.7(2) &  8.3(1) & 40.9(5)\\
 $Q$(6)  & 0.87 & 64.3(1) & 12.5(1) & 11.5(1) & 56.6(5)\\
 $P$(2)  & 0.15 & 63.5(3) & 12.7(8) &  2.0(1) & 49.8(25)\\
$ P$(4)  & 0.34 & 63.8(1) & 12.5(3) &  4.5(1) & 67.8(15)\\
 $P$(6)  & 0.39 & 63.4(1) & 12.5(3) &  5.2(1) & 68.5(13)\\
 $P$(8)  & 0.31 & 64.4(1) & 13.0(4) &  4.2(1) & 52.1(12)\\ 
$ P$(10) & 0.24 & 65.7(2) & 11.7(4) &  2.9(1) & 34.9(12)\\
 $P$(12) & 0.15 & 65.5(4) & 13.0(8) &  2.1(1) & 24.8(12)\\
\hline \vspace{-0.2cm}\\ 
\multicolumn{2}{l}{C$^{13}$CC$^\ast$}  &&&&\\
\vspace{-0.3cm}\\ 
$ Q$(2) & 0.046 & 62.0(3) & 10.8(7) & 0.53(3) & 2.64(15)\\
 $Q$(4) & 0.040 & 62.0(4) & 10.8(10) & 0.46(4) & 2.28(20)\\
\hline \vspace{-0.2cm}\\ 
\multicolumn{2}{l}{$^{13}$CCC}  &&&&\\
\vspace{-0.3cm}\\ 
 $Q$(2) & 0.027 & 62.8(8) & 11.1(15) & 0.32(4) & 1.58(20)\\
 $Q$(4) & 0.021 & 62.9(3) & 11.6(6) & 0.26(1) & 1.26(5)\\
\hline
\hline 
\end{tabular}

$^\ast$ Only H-polarization data was used
\end{table}

\begin{figure}[h]
\begin{center}
\includegraphics[width=0.49\textwidth]{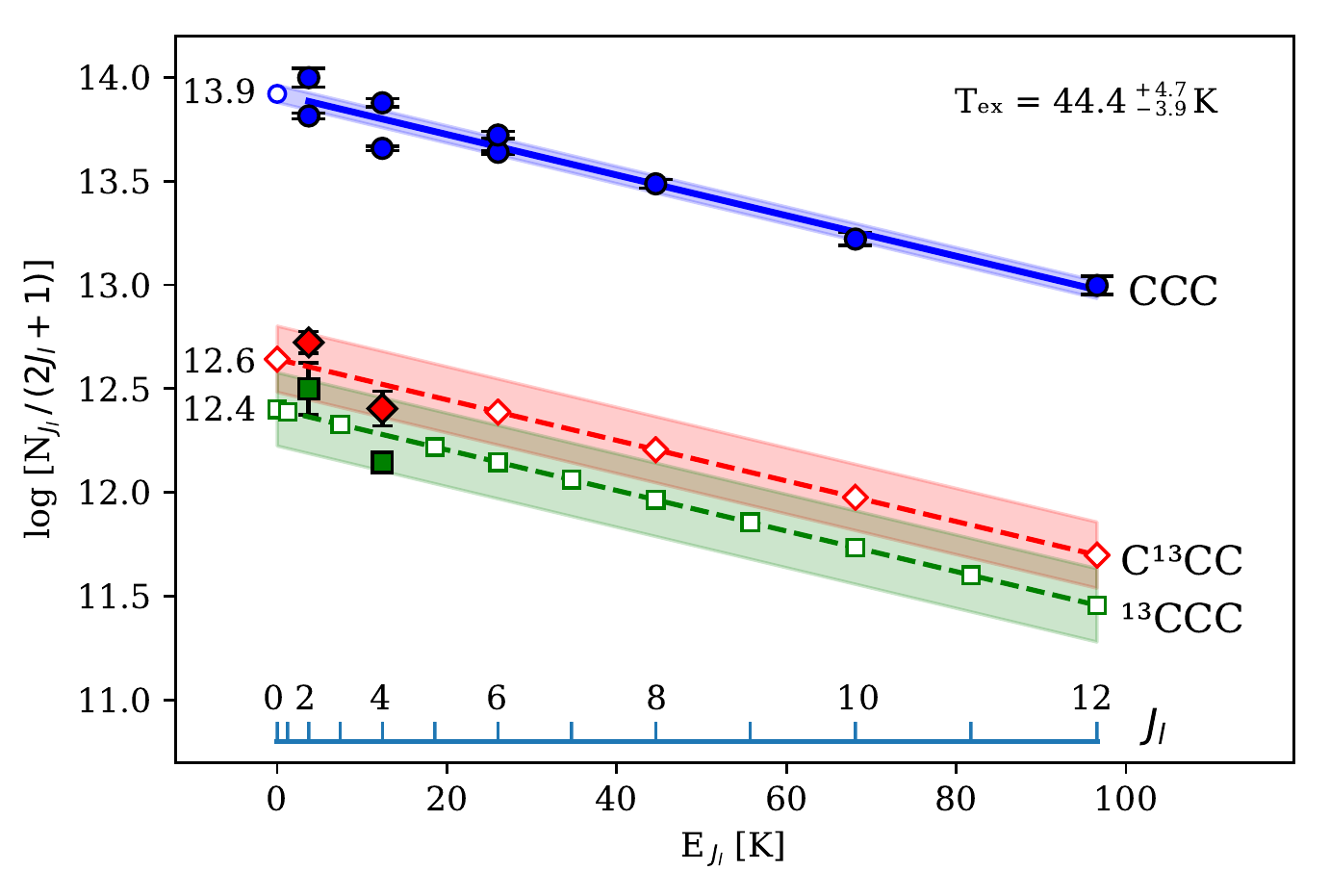}
\caption{Rotation diagram for the \cthree\ lines observed with HIFI/Herschel towards SgrB2(M). The excitation temperature,  $T_{\rm ex}$=44.4$^{+4.7}_{-3.9}$\, K, was derived from a linear fit to the state-specific column densities N$_l$ ($J_l= 2, 4, ... 12$) of the main isotopologue, CCC (blue solid line and blue shaded $\pm 1\sigma$ error range). A straight line with the temperature slope of the main isotopologue was fitted to the data $J_l=2, 4$ of the isotopologues \CthCC\, (red) and \thCCC\, (green). Filled symbols represent measured values, empty symbols are obtained from the least-squares fit. Note, the centro-symmetric species, CCC and \CthCC, have only even numbered $J_l$ levels. The numbers to the left are the $\log ( N_{0})$ values for $J_l=0$ obtained from the fit.
\label{fig_c3rotdiag}}
\end{center}
\end{figure}

The integrated optical depths of the ro-vibrational transitions were used to derive the state-specific column densities $N_{J_l}$ of the lower state rotational levels $J_l$. We used Eqs. (30, 6, 11) from \citet{mangum2015} to rewrite the column density  with respect to the lower state as:

\begin{equationarray}{c}
N_{J_l} = \frac{8\pi \nu^3}{c^3}\frac{g_l}{A_{ul}\,g_u}\left[1-{\rm exp}\left(-\frac{h\nu}{k_BT_{ex}}\right)\right]^{-1}\int{\tau d\upsilon}
\end{equationarray}

where $g_l$ and $g_u$ are the rotational degeneracy factors 2$J+1$ of
the $J_l$ lower and $J_u$ upper rotational level respectively. 
Figure \ref{fig_c3rotdiag} shows the state-specific column densities
$N_{J_l}$ of CCC, \thCCC, and \CthCC\ as a Boltzmann plot, where we used
the CCC column densities to derive the gas excitation temperature
$T_{\rm ex}$. Since the state-specific column density itself is a
function of temperature (Eq.\,2) we used an iterative fitting
procedure until the temperature converged to the final value   of
$T_{ex}$= 44.4$^{+4.7}_{-3.9}$\,K . Note that for $J_l$ = 2, 4, 6  of
CCC, where both, $P$($J$) and $Q$($J$) transitions have been observed, the
column densities of $P$-branch lines are consistently higher than those of
the $Q$-branch lines, which significantly contributes to the obtained
uncertainties  of $T_{ex}$. A similar trend has been seen for previous
CCC observations of other sources \citep{mookerjea2010, mookerjea2012,
mookerjea2014}, however the reason behind this is not clear.

In Figure\,\ref{fig_c3rotdiag} we also present the state-specific
column densities $\log(\rm N_0)$ of $J_l$ = 0 rotational states. The
values for \CthCC\, and \thCCC\, were obtained from a Boltzmann fit to
$Q$(2) and $Q$(4) and under the assumption that the excitation temperatures
of \CthCC\, and \thCCC, and thus the slopes of the Boltzmann plots, are
the same as for CCC.  
Within limits of uncertainties shown by the shaded zone in
Figure\,\ref{fig_c3rotdiag}, the state-specific column densities of
\CthCC\ are found to be systematically larger than those of \thCCC.
This can also be seen from Fig.\,\ref{fig_13c3specs} and
Figure\,\ref{fig_c13ccspecs}  where line intensities of \CthCC\, are
stronger than those of \thCCC. 
 
We calculated the  total column densities $N$ of the three
C$_3$-isotopologues  by using the partition functions Q$_{rv}$ given in
Eq.\,(3). Due to  spin statistics half the rotational levels of the
symmetric molecules CCC and C$^{13}$CC are missing, whereas for the
asymmetric \thCCC\, all rotational states are present. Hence, the
partition function $Q_{rv}$ of the asymmetric molecule  \thCCC\, is
twice as large as those of the symmetric molecules (Table\,\ref{tab_results}).  At 44~K the lowest bending states $v_2$ = 1
($E_1$=90.6~K) and  $v_2$ = 2  ($E_2$=191.4~K) are thermally populated
and the contributions of   $v_2$ = 0, 1, 2 to the partition functions
are  77\%, 19\% and 4\%, respectively. Therefore, in Eq.\,(3) we took
vibrational excitation up to $v_2$ = 2 into account but ignored higher
vibrational levels. The ro-vibrational energy levels $E_{rv}$  were
calculated using the  molecular parameters of \cite{Breier2016},  and
with $g_v$ = 1, 2, 3, the degeneracy factors of the bending states $v_2$
= 0, 1, 2, respectively.

 \begin{equationarray}{c}
Q_{rv}=\sum_{J_l (\rm v=0,1,2)}
(2J_l+1) \, g_v \exp \left(\frac{-E_{rv}(J_l, \rm v)}{k_{\rm B} T_{\rm ex}} \right) .
\end{equationarray}

We used Eq.\,(4) to calculate the total column densities of CCC, \CthCC\, and \thCCC, the results of which are presented in Table~\ref{tab_results}.

\begin{equationarray}{c}
N = Q_{\rm rv} \,\frac{N_{J_l}}{2J_l+1}\, {\rm exp}\left(\frac{E_{rot}(J_l)}{k_B T_{ex}} \right) 
\end{equationarray}

\begin{table}
\centering
\caption{Total column densities and partition functions of CCC,  C$^{13}$CC and  $^{13}$CCC transitions.
\label{tab_results}}
\begin{tabular}{rlcccr}
\hline
\hline\\
  \  & $T_{\rm ex}$(K)  &$ \log\,(N_0)$ &  $Q_{rv}$  & $N \times 10^{14}$ & rel. \\
  \  &    &  &   &  cm$^{-2}$  \, & \ \\
\hline
&&&&\\\vspace{0.2cm}
            CCC & 44.4$^{+4.7}_{-3.9}$& 13.92(04) & 46.64 & $38.8^{+3.9}_{-3.5}$   & 100 \% \\ \vspace{0.2cm}
C$^{13}$CC & 44.4\,$^{\rm a}$ & 12.64(16)$^{\rm b}$ & 47.66 & $2.1^{+0.9}_{-0.6}$ $^{\rm b}$ & 5\,$\pm$2\,\%\\ 
$^{13}$CCC & 44.4\,$^{\rm a}$ & 12.40(18)$^{\rm b}$ & 97.30 & $2.4^{+1.2}_{-0.8}$ $^{\rm b}$   & 6\,$\pm$3\,\% \\ 
\vspace{-0.2cm}\\
 \hline
\hline\\
\end{tabular}
\raggedright
{\small \\
	$^{\rm a}  T_{ex}$ is fixed to the value derived for CCC.\\
	$^{\rm b}$ Uncertainties are given as mean absolute deviation (MAD) of the two measured $Q$(2) and $Q$(4) column densities from their mean value.\\
}
\end{table}

For the main isotopologue we obtained a total column density of $N$(CCC)
= $38.8^{+3.9}_{-3.5}\times 10^{14}$cm$^{-2}$,  the uncertainty of
which is mainly due to the uncertainty of the excitation temperature
$T_{ex}$. The CCC column density is in agreement with the value derived
by \citet{polehampton2007} based on ISO/LWS observations.  For the two
$^{13}$C-isotopologues we derived the column densities  $N$(\CthCC)
=$2.1^{+0.9}_{-0.6}\times 10^{14}$cm$^{-2}$  and $N$(\thCCC) =
2.4$^{+1.2}_{-0.8}\times 10^{14}$ cm$^{-2}$. The single state column densities of $Q$(2) and $Q$(4) deviate by more than the $\pm1\sigma$ error bars from the 44\,K temperature curve (Fig.\,\ref{fig_c3rotdiag}). To give a more realistic error for the total column
densities $N$(\thCCC) and $N$(\CthCC) we used the mean absolute
deviation (MAD) of the $Q(2)$ and $Q(4)$ column densities as
uncertainties of the \thCCC\ and \CthCC\ total column densities. With
regard to the main isotopologue CCC the relative abundances of \CthCC\
and \thCCC\ are  5~\% and 6~\% respectively, which both have large
uncertainties  (see Table~\ref{tab_results}).

\section{Discussion}
 In the previous section we have given total column densities of the
isotopologues CCC, \thCCC\ and \CthCC\ from which the relative abundances  $N(a)$\,/\,$N(b)$, with $a, b$
representing two of the three observed isotopologues, can be derived. Furthermore, we aim to
obtain a value for the $^{12}$C\,/$^{13}$C atomic carbon ratio present in
C$_3$ molecules of the Galactic center, which then can be compared to
results obtained for other carbon containing molecules and their
mono-substituted $^{13}$C isotopologues. The $N(a)$\,/\,$N(b)$ ratios can be
calculated directly from the total column densities $N$ listed in Table
\ref{tab_results}, and the $^{12}$C\,/$^{13}$C ratio can be derived from:
  
\begin{equationarray}{c}
{\rm
\frac{^{12}C}{^{13}C} = \frac{3\,{\it N}(CCC)}{{\it N}(^{13}CCC) + {\it N}(C^{13}CC)}
}
\end{equationarray}

The ratios $N(a)$\,/\,$N(b)$ obtained from the total column densities are
listed in Table \ref{tab_peaks} as  $N(a)$\,/\,$N(b)$ TOTAL, and the
corresponding  $^{12}$C\,/\,$^{13}$C ratio  is given at the end of the row.
The large errors of up to 50\% are mainly due to the uncertainties of the \thCCC\ and \CthCC\ total column densities
given in Table~\ref{tab_results}. As an alternative method, the $N(a)$\,/\,$N(b)$ ratios can be calculated from single state-specific column densities $N^a_{J_l}$
and $N^b_{J_l}$ of the isotopologues $a$ and $b$ via Eq. (4):  

\begin{equationarray}{c}
{
\frac{N(a)}{N(b)} = \frac{Q_{rv}^a \, N^a_{J_l} \, (2J_l^b+1)} {Q_{rv}^b \, N^b_{J_l}\,(2J_l^a+1)} \cdot {\rm exp}\left(\frac{\Delta E_{rot}}{k_B T_{ex}} \right) 
}
\end{equationarray}  
with the energy difference $\Delta E_{rot}$ of the rotational levels $J_l^a$ and $J_l^b$, which value is close to zero if we choose $J_l$ for both isotopologues the same. In our analysis we calculated the ratios for $Q(2)$ and $Q(4)$ separately, using the state-specific column densities given in Table~\ref{tab_obsdata}. Similar to Eq. (5) the $^{12}$C\,/$^{13}$C isotope ratio of C$_3$ can be calculated from Eq. (7), neglecting the small energy differences between rotational levels of identical quantum numbers $J_l$ of the three isotopologues:
\begin{equationarray}{c}
{
\frac{^{12}C}{^{13}C}\approx  \frac{3\,Q_{\rm rv}^{\rm CCC}\,{\it N}_{J_l}(\rm CCC)}{\,Q_{\rm rv}^{^{13}\rm CCC}\,{\it N}_{J_l}(^{13}\rm CCC) + \,Q_{\rm rv}^{\rm C^{13}CC}\,{ N}_{J_l}(\rm C^{13}CC)}
}
\end{equationarray}  

The ratios $N(a)$\,/\,$N(b)$ derived from state-specific column densities of
$Q(2)$ and $Q(4)$ via Eq. (6) as well as the corresponding
$^{12}$C\,/\,$^{13}$C ratios calculated from Eq. (7) are listed in Table
\ref{tab_peaks}.  Note, the given uncertainties are due to the small uncertainties of the $Q(2)$ and $Q(4)$ state specific column densities which do not depend on the temperature $T_{ex}$. Nevertheless, the ratios derived from $Q(2)$ and $Q(4)$ should agree within error bars if thermal equilibrium is assumed. We found that the results for Q(2) and Q(4) differ significantly and concluded that their uncertainties are largely underestimated. Therefore, in Table \ref{tab_peaks} we give the mean values of Q(2) and Q(4) as  $N(a)$\,/\,$N(b)$ AV and  their mean absolute deviations (MAD) in brackets. Table\,\ref{tab_peaks} also presents the expected values $N(a)$\,/\,$N(b)$ (EXPEC) for an assumed $^{12}$C\,/$^{13}$C ratio of 20, the value typically found for gas near the Galactic center.\\

\begin{table}	
	\centering
	\caption{ Derived $^{12}$C/$^{13}$C ratios in C$_3$ and fractional abundances of C$_3$ isotopologues. The $\pm1\sigma$ standard deviations are given in brackets.  
		\label{tab_peaks}}
	\begin{tabular}{lcccc}
\toprule
\hline\vspace{-0.2cm}\\
	${N^a}/{N^b}$ 		    
	&  \large $\genfrac{}{}{0.5pt}{1}{N(\rm CCC)}{N(\rm C^{13}CC)}$  
	&  \large $\genfrac{}{}{0.5pt}{1}{N(\rm CCC)}{N(\rm ^{13}CCC)}$   
	&  \large $\genfrac{}{}{0.5pt}{1}{N(\rm ^{13}CCC)}{N(\rm C^{13}CC)}$ 		&  \large $\genfrac{}{}{0.5pt}{1}{^{12}\rm C}{^{13}\rm C}$  \\ 	
\vspace{-0.2cm}\\
 \hline \vspace{-0.1cm} \\
  TOTAL      &   18.6(7.0)       &  16.0(6.7)             &  1.2(0.6)         &  25.8(7.5)  \\   
     Q(2)     &   12.1(0.7)        &   9.9(1.3)              &  1.2(0.2)         &  16.3(1.2)  \\  
     Q(4)     &   17.6(1.6)        & 15.6(0.6)               &  1.1(0.1)         &  24.7(1.2)  \\   
\vspace{-0.2cm}\\
\hline \vspace{-0.2cm}\\
    AV         &   14.8(2.7)$^{\dagger}$         & 12.7(2.8)$^{\dagger}$                 & 1.2(0.1)$^{\dagger}$         &  20.5(4.2)$^{\dagger}$  \\ 
  EXPEC      &    20                &   10                     &     2               &   20             \\  
 		\bottomrule\addlinespace[5pt]
	\end{tabular}
	{\\
		\raggedright
		\small{
			TOTAL: Calculated from total column densities $N$ given in Tab.\ref{tab_results}.\\  
			Q(2), Q(4): Calculated via Eq.(6) from N$_{J_l}$ values in Tab.\ref{tab_obsdata}.\\
			AV $^{\dagger}$: Mean of Q(2) and Q(4) and mean absolute deviations (MAD).\\
EXPEC: Statistically expected values for $^{12}$C\,/\,$^{13}$C=20.\\ 
		}   
	}
\end{table}

We conclude that the derived  averaged (AV) $^{12}$C/$^{13}$C ratio
of 20.5$\pm$4.2 is in good agreement with the statistically expected
ratio of 20. The averaged value
(AV) $N$(CCC)\,/\,$N$($^{13}$CCC)=12.7$\pm$2.8 is larger than the expected
value of 10 but both values agree within error bars. In opposite,  the
ratio $N$(CCC)\,/\,$N$(C$^{13}$CC)=14.8$\pm$2.7 is significantly smaller
than the expected value of 20, which indicates a moderate enhancement of
the symmetric species \CthCC . Accordingly, the ratio
$N$($^{13}$CCC)\,/\,$N$(C$^{13}$CC)=1.2$\pm$0.1 is smaller than the
statistically expected value of 2. This isotopic shift in favor of the
centrally substituted species \CthCC\, can be explained either by a less
effective formation of \thCCC\, or by isotope position-exchange
reactions. In Eq.\,(8) the reactant X($^{12}$C) can be, e.g. C$^+$, or
any carbon molecule or molecular ion that exchanges a $^{12}$C carbon
atom and changes the ordering of carbon atoms in the C$_3$ molecule. 

\begin{equationarray}{c}
{\rm
X(^{12}C)\,+\, ^{13}CCC\,  \rightleftharpoons \, C^{13}CC \,+\, X(^{12}C) \,+\, \Delta{\it E_0}
}
\end{equationarray}     

For Eq.\,(8) the equilibrium is determined by the constant $k_p$ of a
chemical reaction:

\begin{equationarray}{c}
k_p= \exp \left(\frac{-\Delta G}{k_B\,T_{\rm ex}}\right)=\frac{Q\,^{^{13}\rm CCC}}{Q\,^{\rm C^{13}CC}}\exp\left(\frac{-\Delta E_0}{k_B\,T_{\rm ex}}\right) 
\end{equationarray}     

Here, $\Delta G$ and $\Delta E_0$ are the differences in Gibbs energy
and the zero point energy, respectively.  We used calculated vibrational
energies of the C$_3$-isotopologues published by \citet{Schroeder2016}
to calculate accurate zero point energies. We found that \CthCC\ has a
zero point energy smaller by 15.9\,K than \thCCC\ and hence \CthCC\ is
more stable than \thCCC.  At high excitation temperatures the reaction
constant $k_p$ converges to the ratio of the partition functions which
equals 2 (as obtained from a statistical distribution of isotopes). On
the other hand, at an excitation temperature of 44~K the value of $k_p$
is reduced by a factor of 0.7, which leads to an expected abundance
ratio of $N$(\thCCC)/$N$(\CthCC) = 1.4, a result which is in good
agreement with the observed value of 1.2$\pm$1. Note, the equilibrium of
Eq.\,(8) only depends on the amount of $^{12}$C but is independent of
the available $^{13}$C budget. 

Furthermore, as has extensively been studied in case of
ozone,\,O$_3$,\,the reaction rate that leads to the formation of the
symmetric isotopologue  can significantly differ from the reaction rate
that forms the asymmetric species, see e.g. \citep{Feilberg2013}. A
review of mass-independent isotope effects in chemical reactions was
published by \citet{Thiemens2006}. In general, the symmetry effect in
rate coefficients depends on the spin statistical weights, as was
explained in case of ozone by \citet{Gellene1996}. It is largest for
nuclei of zero spin, when half of the molecular states are missing due
to spin statistical reasons, as is the case for $^{16}$O in reaction
with $^{18}$O.  The symmetry effect may also be relevant for $^{12}$C in
reaction with $^{13}$C, but to our knowledge this has not yet been
considered in the formation process of C$_3$.

Till now, the abundance of \cthree\ in the warm envelopes of hot cores
have been consistently explained in terms of chemical pathways involving
post-warm-up gas phase chemistry of CH$_4$ released from grain surface
\citep{mookerjea2012}. This warm-carbon-chain-chemistry (WCCC) network
also forms molecules like CCH, $c$-C$_3$H$_2$, CH$_3$CCH, HC$_3$N,
CH$_3$CN etc.  \citep{mookerjea2012}.  Based on a spectral line survey
with IRAM\,30m telescope \citet{belloche2013} derived the column
densities of CO and several hydrocarbons including,  e.g., CCH,
CH$_3$CCH, $c$-C$_3$H$_2$, CH$_3$CN HC$_3$N and their multiple $^{13}$C
isotopologues. In this analysis, the authors derived a multi-component
global fit to the spectrum assuming local thermodynamic equilibrium
(LTE) and found that for all these species, including CO the
$^{12}$C/$^{13}$C ratio is $\sim 20$. For the molecules CCH, CH$_3$CCH,
$c$-C$_3$H$_2$, CH$_3$CN and HC$_3$N the $^{12}$C/$^{13}$C ratio is 20,
17, 20, 12 and 13, respectively, which within limits of uncertainties is
consistent with our finding based on \cthree. However, in contrast to
our findings for \cthree\ and its isotopologues, the isotopic ratio in all
these molecules in SgrB2(M) is identical irrespective of the position of
the isotope-substituted carbon atom in the molecule. 

The non-equivalence of
$^{13}$C-substitution in carbon chain molecules have also been observed in
dark clouds, low-mass star forming regions as well as in hot cores.
\citet{sakai2010} determined the [C$^{13}$CH]/[$^{13}$CCH] ratio for the
dark cloud TMC-1 and the low-mass star forming core L1527 to be $\sim
1.6$ and \citet{taniguchi2016a} observed all three  $^{13}$C isotopologues of
HC$_3$N in the hot core G28.28-0.36 and found that HCC$^{13}$CN is more
abundant by a factor of 1.4 compared to the other two equally abundant
isotopologues.  These authors explained the difference in abundances of the
different isotopomers in terms of the formation pathway involving
neutral-neutral reaction for the respective molecular species.
Additionally, \citet{furuya2011} pointed out that because of the
difference in zero-point vibrational energy of the end-substituted and
centrally-substituted species like CCH and CCS, it is also possible to
have an exchange of the $^{13}$C position by isotopomer-exchange
reaction.

\citet{yoshida2015} studied the isotopic abundances of $c$-C$_3$H$_2$ in
L1527 and found for $c$-C$_3$H$_2$ that the symmetric species
$c$-CC$^{13}$CH$_2$ with a lower zero-point energy has a larger
abundance than the asymmetric  $c$-C$^{13}$CCH$_2$ species.
\citet{yoshida2015} suggested position-exchange reactions  as a possible
mechanism of the relative enrichment of one of the isotopologues.  Among all
these molecules \cthree\, is closest to $c$-C$_3$H$_2$ since both
molecules contain three carbon atoms of which two are equivalent and are
formed in dense clouds via WCCC, and  furthermore, the two molecules
show similar trends in abundances of $^{12}$C and $^{13}$C species.

\section{Summary}

We have presented the first detection of the singly-substituted
$^{13}$C-isotopologues \thCCC\ and \CthCC\ along with nine
ro-vibrational transitions of the main isotopologue, CCC, towards the
high mass star forming core SgrB2(M).  All the transitions are detected
in absorption and around velocities of 62--64\,\kms, suggesting that the
absorption features are due to warm molecular gas physically associated
with the background source. We estimated a rotational temperature of
44.4\,K to  explain the optical depths of all the observed transitions
of \cthree\ assuming LTE. The column densities for CCC, \thCCC\ and
\CthCC\ were estimated considering the excitation temperatures of all
the species to be identical and assuming LTE. We find that the
$^{12}$C/$^{13}$C abundance ratio in C$_{3}$ is 20.5$\pm 4.2$,
which is in agreement with the observed atomic ratio of 20 in SgrB2(M).
We find the $N$(\thCCC) / $N$(\CthCC) ratio to be 1.2$\pm$0.1 as
opposed to the statistically expected value of 2 and propose that this
discrepancy arises due to the lower zero-point energy of \CthCC\ which
makes position-exchange reaction converting \thCCC\ to \CthCC\
energetically favorable.

\begin{acknowledgements}

GREAT is a development by the MPI f\"ur Radioastronomie and
KOSMA/Universit{\"a}t zu K\"oln, in cooperation with the DLR Institut
f\"ur Optische Sensorsysteme. The development ofGREAT is financed by the
participating institutes, by the German Aerospace Center (DLR) under
grants 50 OK 1102, 1103 and 1104, and within the Collaborative Research
Centre 956, funded by the Deutsche Forschungsgemeinschaft (DFG). 
T.F.~Giesen was supported by project B2 within CRC~956; A.A.~Breier and 
T.F.~Giesen were supported by DFG SPP-ISM 1573. The
work by R.~Simon and J.~Stutzki was supported by project A4 within
CRC~956; B.~Mookerjea received travel support as a visiting scientist
from CRC~956. 

SOFIA is jointly operated by the Universities Space Research
Association, Inc. (USRA), under NASA contract NAS2-97001, and the
Deutsches SOFIA Institut (DSI) under DLR contract 50 OK 0901 and 50 OK
1301 to the University of Stuttgart. We thank the SOFIA operations and
engineering teams for their dedication and supportive response.

\end{acknowledgements}
 
{} 
\end{document}